\documentclass[fleqn,twoside,twocolumn,nofootinbib,showkeys]{revtex4} 
\usepackage[sec,nocpr]{ujp} 

\begin{document}
\title[On Pendry's Effective Electron Mass]
{ON PENDRY'S EFFECTIVE ELECTRON MASS}
\author{V.V.~Gozhenko}
\affiliation{Institute of Physics, Nat. Acad. of Sci. of Ukraine}
\address{46, Nauky Ave., Kyiv 03680, Ukraine}
\email{vigo@iop.kiev.ua}
 \udk{???} \pacs{45.20.Jj, 45.50.-j,\\[-3pt] 78.67.Pt} \razd{\secvii}

\autorcol{V.V.\hspace*{0.7mm}Gozhenko}

\setcounter{page}{1013}%

\begin{abstract}
In 1996, J.\,\,Pendry, an English theoretical physicist put forward
an idea about the dependence of the effective electron mass on the
magnetic field, while interpreting the dielectric response of metal
wire mesh structures.\,\,The idea was based on the well-known
relation between the kinematic and canonical momenta of a charged
particle moving in the magnetic field.\,\,In this paper, proceeding
from the universal character of that relation, the applicability of
Pendry's effective electron mass $m_{\text{eff}}$ to the problem of
electrons in metal mesh structures, as well as to a wide class of
problems for charges moving in the magnetic field, has been
examined.\,\,The general properties of $m_{\text{eff}}$ following
directly from its definition were found, and an analogy between the
effective electron mass $m_{\text{eff}}$ and $m^{\ast}$ known in the
solid-state theory was established.\,\,A physical interpretation of
$m_{\text{eff}}$ was proposed.\,\,It was demonstrated in several
examples that, despite the generality of the defining relation for
the effective mass $m_{\text{eff}}$, the use of $m_{\text{eff}}$
beyond the problem of the dielectric response of metal wire mesh
structures leads to incorrect results.
\end{abstract}
 \maketitle

\section{Introduction}

The concept of effective mass for charge carriers is widely used to
describe various phenomena in metals and semiconductors.\,\,In
particular, it is used in the theories of electroconductivity, Hall
effect, and cyclotron resonance in metals
\cite{Kittel,Ashcroft,Davydov}.

In the solid-state theory, the effective mass of conduction
electrons, $m^{\ast}$, is usually determined from their dispersion
law $E(\mathbf{k})$, where $E$ and $\mathbf{k}$ are the energy and
the wave vector of an electron, respectively. For example, in the
the simplest case of isotropic quadratic dispersion law
\cite{Davydov}, we have
\[
m^{\ast}\equiv\left( \! \frac{1}{\hbar^{2}}\frac{\partial^{2}E}{\partial k^{2}%
}\!\right)  ^{\!\!-1}\!.
\]
The difference between $m^{\ast}$ and the rest mass of a free
electron, $m$, is known to emerge owing to the interaction of
conduction electrons in a solid with the periodic internal crystal
field.\,\,As a result of this interaction, electrons in the solid
are accelerated under the action of external electromagnetic fields,
however, not as free particles with the mass $m$, but as certain
imaginary particles with the mass $m^{\ast}$.\,\,According to
experimental data, the value of $m^{\ast}$ can significantly (by
1--2 orders of magnitude) differ from $m$ \cite{Kikoin}.

In the case of anisotropic dispersion, the effective mass of the
electron depends on the direction of its acceleration.\,\,In this
case, the dynamical properties of the electron are not characterized
by a single scalar $m^{\ast}$, but a collection of three scalars
$m_{i}^{\ast}$ ($i=x,y,z$), which are reciprocals of the principal
values of the tensor of inverse effective mass of
electrons,~\cite{Davydov}
\[
\left(  m^{\ast}\right)  _{ij}^{-1}=\left(\!
\frac{1}{\hbar^{2}}\frac {\partial^{2}E}{\partial k_{i}\partial
k_{j}}\!\right)\!\!.
\]
It is worth noting that the effective mass $m^{\ast}$ (or
$m_{i}^{\ast}$) is determined by specific features in the electron
energy spectrum of that or another solid, being a function of the
wave vector $\mathbf{k}$, temperature, and
pressure~\cite{Tsidilkovsky}.

In 1996, Pendry \textit{et al.} \cite{Pendry96}, while considering
the frequency dependence of the dielectric function of metal wire
mesh structures, introduced the concept of effective electron mass
dependent on the magnetic field.\,\,Unlike the effective mass of
electrons $m^{\ast}$, which is considered in the solid-state theory,
the effective mass $m_{\text{eff}}$ introduced in work
\cite{Pendry96} (below, it will be called Pendry's effective mass)
is defined on the basis of the relation, known from the classical
mechanics, between the kinematic and canonical momenta of a
non-relativistic spinless charged particle moving in the magnetic
field (similarly to works
\cite{Pendry96,Ramakrishna,Cai}, the SI units are used below):%
\begin{gather}
m_{\text{eff}}\boldsymbol{\mathbf{v}}\equiv\mathbf{P},\label{eq:1}\\
\mathbf{P}=\mathbf{p}+q\mathbf{A},\label{eq:2}%
\end{gather}
where $\mathbf{p}\equiv m\mathbf{v}$ and $\mathbf{P}\equiv\partial
L/\partial\mathbf{v}$ are the kinematic and canonical, respectively,
momenta of the particle; $m$, $\mathbf{v}$, $q$, and $L$ are its
mass, velocity, and charge and the Lagrangian, respectively; and
$\mathbf{A}$ is the vector potential of the magnetic field
($\mathbf{B}=\mathrm{rot}\,\mathbf{A}$).\,\,Using the idea of
$m_{\text{eff}}$, Pendry \textit{et al.} derived an analytical
expression for the mesh plasma frequency $\omega_{p}$, which is in
good agreement with experimental data (see, e.g.,
works~\cite{Pendry98,Gozhenko}).

Pendry's effective mass $m_{\text{eff}}$ cardinally differs from the
mass $m^{\ast}$ considered in the solid-state theory: not only by
definition, but also by properties.\,\,For instance, in contrast to
$m^{\ast}$, the value of $m_{\text{eff}}$ can easily reach giant
values of the order of $( 10^{4}$$\div$$10^{6}) m$
\cite{Pendry96,Ramakrishna}.\,\,Moreover, as a result of the gage
invariance of the vector potential $\mathbf{A}$ entering
Eq.\,(\ref{eq:2}), $m_{\text{eff}}$ turns out an ambiguously defined
quantity \cite{Walser}.\,\,Taking this fact into account, the
critical remarks were made concerning both the correctness of the
definition of $m_{\text{eff}}$ \cite{Walser} and the reality of
predictions on the basis of $m_{\text{eff}}$
\cite{Mikhailov,Solymar}.\,\,Those remarks forced some authors to
search for the ways of deriving the major results of work
\cite{Pendry96} without attracting the $m_{\text{eff}}$
concept.\,\,The success attained in those searches
\cite{Sarychev,Maslovski,Marques,Nefedov} testifies that, despite
the productivity of this concept demonstrated by Pendry \textit{et
al., }its introduction is not required for the explanation of the
dielectric response of a metal mesh structure.\,\,Nevertheless, in
last books on metamaterials \cite{Ramakrishna,Cai}, this response
was explained, by engaging $m_{\text{eff}}$.

Can the notion of $m_{\text{eff}}$ be used beyond the scope of work
\cite{Pendry96}? Relation (\ref{eq:2}) between $\mathbf{p}$ and
$\mathbf{P}$ is rather general, being used not only in classical
mechanics, but also in quantum one.\,\,Proceeding from its universal
character and on the basis of the principle of integrity of physics,
one may hope for that the concept of $m_{\text{eff}}$ could be
applied, while considering a wide range of problems (both classical
and quantum-mechanical) concerning the motion of charges in a
magnetic field.\,\,However, the possibility of its application has
not been considered till now.

Earlier, the interpretation of $m_{\text{eff}}$ was not
considered.\,\,The absence of such an interpretation that would be
compatible at least with the results of work \cite{Pendry96} invokes
a number of questions concerning the physical meaning of the
effective mass $m_{\text{eff}}$.\,\,For example, does
$m_{\text{eff}}$ characterize the inertial properties of an electron
in the magnetic field?\,\,Or does the relation
$m_{\text{eff}}=10^{4}m$ mean that the inertial and, owing to the
Einstein equivalence principle, gravitational masses of an electron
increase in the magnetic field by four orders of
magnitude?\,\,Statements like \textquotedblleft Electrons become as
heavy as hydrogen atoms\textquotedblright\ \cite[p.~4775]{Pendry96}
are based on a free interpretation of the relation
$m_{\text{eff}}=10^{4}m$ rather than the physical interpretation of
$m_{\text{eff}}$ itself, and more likely entangle the situation
rather than clarify it.

It is worth to recall that analogous questions concerning the
\textquotedblleft ordinary\textquotedblright\ effective mass
$m^{\ast}$ of electrons in a solid arose when the solid-state theory
was formulated in the 1930s, in particular, during the discussions
concerning the interpretation of the results of classical
experiments by Tolman and Stewart \cite{Tolman} and Barnett
\cite{Barnett}.\,\,Answers to them were given only after the proper
interpretation of $m^{\ast}$, which was based on its definition and
properties, had appeared (see, e.g., work \cite{TsidilkovskyUFN}).

The problem concerning the interpretation of the effective mass
$m_{\text{eff}}$ proposed by Pendry \textit{et al.} could also be
removed after a detailed analysis of the $m_{\text{eff}}$ properties
that follow from its definition. However, to the knowledge of the
author of this work, nobody has carried out such an analysis till
now.

\begin{figure*}%
\vskip1mm
\includegraphics[width=12cm]{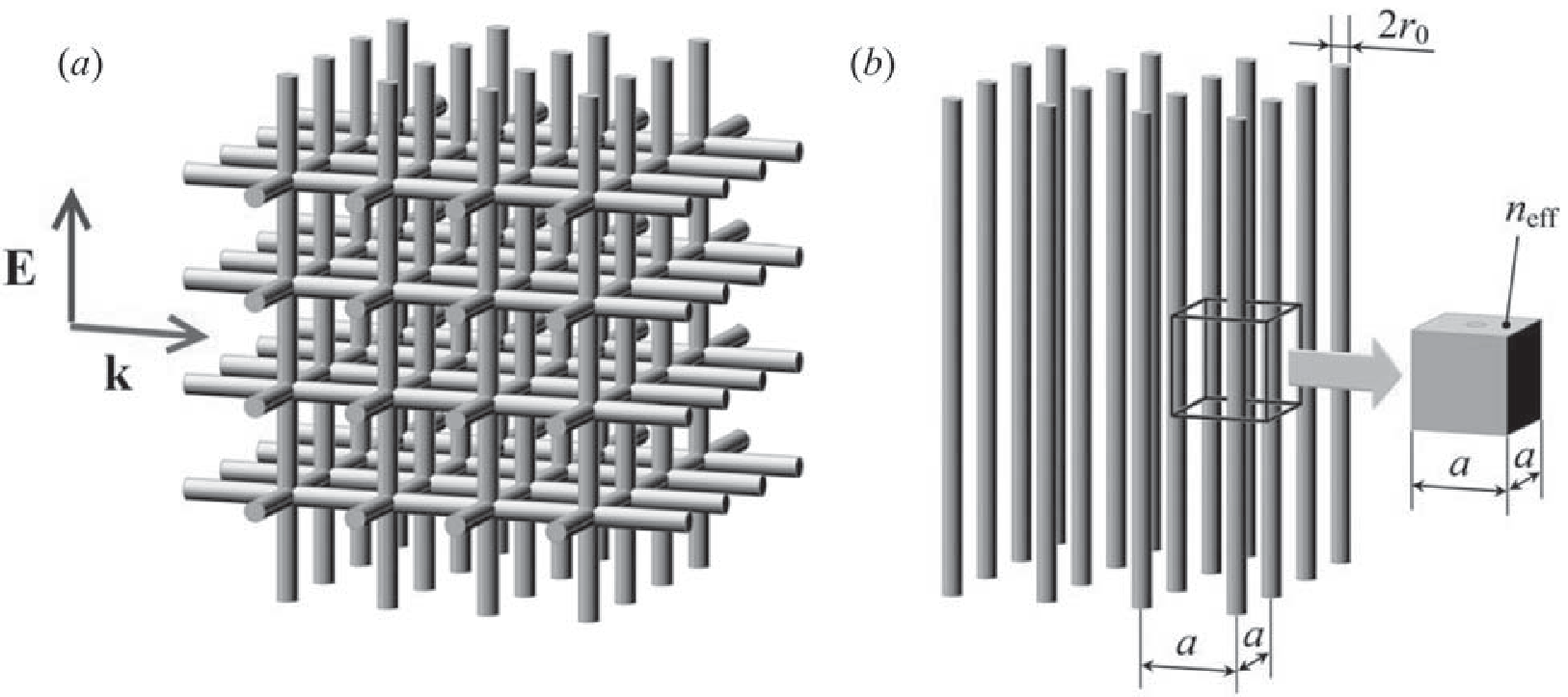}
\vskip-3mm\caption{Scheme of a metal wire mesh structure excited by
an external electromagnetic wave.\,\,Electric oscillations arise in
the mesh conductors (they are called active), which are parallel to
the vector $\mathbf{E}$ of the incident wave. If the orientation of
$\mathbf{E}$ with respect to the mesh structure is arbitrary, all
the conductors, generally speaking, are active ({\it a}).\,\,At the
homogenization of the sublattice of active conductors in the cubic
metal mesh structure, its unit cell, which covers a conductor
segment with an arbitrary length, is considered as being filled with
a homogeneous electron plasma with concentration $n_{\text{eff}}$
({\it b}) }
\end{figure*}

This work was aimed at establishing the general properties of the effective mass
$m_{\text{eff}}$, which follow directly from its definition; formulating a
physical interpretation of $m_{\text{eff}}$, which would be compatible, at
least, with the verified results of work \cite{Pendry96}; and analyzing the
application of $m_{\text{eff}}$ beyond the problem of metal wire mesh
structures, namely, in a wide range of problems concerning the motion of
charges in the external magnetic field, when the relation between $\mathbf{p}$
and $\mathbf{P}$ in the definition of $m_{\text{eff}}$ should manifest itself.

The paper is organized as follows.\,\,In Section 2, the main results
of work \cite{Pendry96} concerning the effective mass
$m_{\text{eff}}$ of electrons in metal mesh structures and the
plasma frequency $\omega_{p}$ of those meshes are presented.\,\,It
was done not only to demonstrate a way, in which the concept of
$m_{\text{eff}}$ is used at the solution of a specific problem, but
also to give the reader unfamiliar with this concept some initial
information on $m_{\text{eff}}$, in particular, its order of
magnitude and its dependence on the parameters of the problem.\,\,In
Section~3, on the basis of definition (\ref{eq:1}), the general
properties of $m_{\text{eff}}$ are found, and an interpretation of
$m_{\text{eff}}$ is proposed.\,\,In Section 4, a number of examples
illustrating the application of the $m_{\text{eff}}$ notion in
various situations associated with the motion of classical or
quantum-mechanical charged particles in a magnetic field are
considered.

\section{Effective Mass of an Electron\\ in Metal Wire Mesh Structures}

The metal wire mesh structures are, as a rule, three-dimensional
periodic structures fabricated from thin metal wire conductors
(Figure,~\textit{a}).\,\,Such meshes have already been studied for
more than half a century: first, as artificial insulators and, at
the modern stage, as photonic crystals and metamaterials (see a
short historical excursion in work \cite{Simovski}).\,\,In the
branch of metamaterials, the wire meshes are more often called
\textit{mesh metamaterials} and usually applied as media with a
negative effective dielectric permittivity
$\varepsilon_{\text{eff}}$. Since the dielectric function
$\varepsilon_{\text{eff}}(\omega)$ of mesh structures has a Drude
form (it was established as early as in the 1960s), negative values
of $\varepsilon_{\text{eff}}$ are attained at frequencies lower than
the plasma frequency $\omega_{p}$ of the mesh.\,\,The plasma
frequency $\omega_{p}$ depends on the mesh structure
parameters.\,\,The knowledge of the explicit form of this dependence
is important both for the mesh design and for the interpretation of
experimental data.\,\,For the simplest model of a cubic mesh
structure composed
of infinitely thin ideal conductors, this dependence looks like%
\begin{equation}
\omega_{p}=\sqrt{\frac{2\pi c^{2}}{a^{2}\ln\left(  a/r_{0}\right)  }%
},\label{eq:3}%
\end{equation}
where $a$ is the mesh period, $r_{0}$ the radius of conductor wires
($r_{0}\ll a$), and $c$ the speed of light in vacuum.\,\,Expression
(\ref{eq:3}) was derived for the first time in work \cite{Pendry96}
on the basis of the following reasoning.

For continuous media, the plasma frequency is known to be given by the
Langmuir formula \cite{Langmuir}
\begin{equation}
\omega_{p}=\sqrt{\frac{ne^{2}}{\varepsilon_{0}m}},\label{eq:4}%
\end{equation}
where $n$ is the concentration of electrons in the medium, $m=9.11\times
10^{-31}$~kg is the electron mass, $e=1.60\times10^{-19}~$C is the elementary
charge, and $\varepsilon_{0}=8.85\times10^{-12}$~F/m is the electric constant.
Formula (\ref{eq:4}) can be elementarily derived, while considering the dynamics
of electrons at plasma oscillations.

The plasma frequency of a cubic metal mesh structure equals that of
the sublattice of its active conductors.\,\,This sublattice can be
regarded as a certain continuous medium (Figure,~\textit{b}) with
the effective electron
concentration%
\begin{equation}
n_{\text{eff}}=\frac{\pi r_{0}^{2}}{a^{2}}n,\label{eq:5}%
\end{equation}
where $n$ is the electron concentration in the material of mesh
conductors. In addition, it was proposed in work \cite{Pendry96} to
take into account that the dynamics of electrons in the mesh is
influenced by the magnetic field of currents $I$, which arise in the
active conductors of the mesh structure at plasma
oscillations.\,\,This influence can be effectively presented as a
change of the electron mass $m\rightarrow m_{\text{eff}}$, where
$m_{\text{eff}}$ is defined by formulas (\ref{eq:1}) and
(\ref{eq:2}), in which $\mathbf{A}$ is the vector potential of the
magnetic field induced by the currents~$I$.

For an infinite mesh structure, the vector potential $\mathbf{A}$ of
currents $I$ is periodic.\,\,In every unit cell of the mesh centered
at the conductor axis, it can be calculated by the formula
\cite{Pendry96,Pendry98}
\begin{equation}
A(r)=\frac{\mu_{0}I}{2\pi}\ln(a/r),\label{eq:6}%
\end{equation}
where $r$ is the distance from the conductor axis ($r\in\lbrack
r_{0},a/2]$), and $\mu_{0}=4\pi\times10^{-7}$~H/m is the magnetic
constant.\,\,Taking into account that $I=\pi r_{0}^{2}nev$, where
$v$ is the drift velocity of electrons in the conductor, and that
electrons move in ideal conductors only on their surface and,
therefore, \textquotedblleft feel\textquotedblright\ the
field $A(r_{0})$, we obtain%
\[
A(r_{0})=\frac{\mu_{0}r_{0}^{2}nev}{2}\ln(a/r_{0}).
\]
For real mesh metamaterials, the term $eA(r_{0})$ in
Eq.~(\ref{eq:2}) turns out much larger than the kinematic momentum
$mv$ of conduction electrons (see the numerical example
below).\,\,Therefore, neglecting the latter, it is possible to
assume that
\[
m_{\text{eff}}=\frac{eA(r_{0})}{v}%
\]
or\vspace*{-3mm}%
\begin{equation}
m_{\text{eff}}=\frac{\mu_{0}r_{0}^{2}ne^{2}}{2}\ln(a/r_{0}).\label{eq:7}%
\end{equation}
This is a sought expression for Pendry's effective electron mass in
metal mesh structures.\,\,Formula (\ref{eq:3}) for the mesh plasma
frequency can now be easily obtained from the Langmuir formula
(\ref{eq:4}) by substituting Eqs.~(\ref{eq:5}) and (\ref{eq:7}) for
$n$ and $m$, respectively, into it.

Below, we give a few remarks concerning $m_{\text{eff}}$.

1.\,\,An idea about the magnitude of effective electron mass in
metal wire mesh structures can be obtained from the following
numerical example
\cite{Pendry96}.\,\,For a mesh with the parameters $a=5$~mm and $r_{0}=1$%
~$\mu\mathrm{m}$ (so that $a/r_{0}=5000\gg1$) and made from aluminum
conductors ($n=1.81\times10^{29}~\mathrm{m}^{-3}\text{)}$, we have%
\[
m_{\text{eff}}/m\approx2.7\times10^{4}.
\]
By the way, the giant value of this ratio ($m_{\text{eff}}/m\gg$
$\gg1$) means that the relation $eA\gg mv$ used in the derivation of
expression (\ref{eq:7}) for $m_{\text{eff}}$ is satisfied.

The electron dilution coefficient for the considered mesh structure equals%
\[
n_{\text{eff}}/n=a^{2}/\left(  \pi r_{0}^{2}\right)  \approx8.0\times10^{6},
\]
and the mesh plasma frequency turns out by five orders of magnitude lower than
the plasma frequency $\omega_{p}($Al$)$ for bulk aluminum,%
\begin{equation}
\frac{\omega_{p}(\text{Al})}{\omega_{p}}=\sqrt{\frac{n}{n_{\text{eff}}}%
\frac{m_{\text{eff}}}{m}}\approx4.6\times10^{5}.\label{eq:8}%
\end{equation}
Estimate (\ref{eq:8}) is confirmed by experimental data.

2.\,\,As one can see from the example given above, the increase of
the electron mass, $m\rightarrow m_{\text{eff}}$, plays a
sub\-stan\-tial role in the formation of the value of
$\omega_{p}$.\,\,The account for only one effect~-- the dilution of
charge car\-riers, $n\rightarrow n_{\text{eff}}$, which is
intuitively clear~-- would not allow one to obtain correct numerical
values
for~$\omega_{p}$.

3.\,\,According to Eq.~(\ref{eq:7}), the effective electron mass
depends on the geometrical mesh parameters $a$ and $r_{0}$ in such a
way that, by increasing the mesh period, i.e.\,\,passing to the
limiting case of isolated conductor, \textquotedblleft one could
easily obtain electrons with an effective mass of, say, 1
kg\textquotedblright\ \cite{Mikhailov}.

4.\,\,The authors of work \cite{Pendry96} assume the mutual
inductance $L$ of mesh's conductors to be the physical origin of the
electron mass increase.\,\,However, while deriving formulas for
$m_{\text{eff}}$ and $\omega_{p}$, the quantity $L$ has not been
used.

\section{Properties and Interpretation of \boldmath$m_{\text{eff}}$}\vspace*{-1mm}

\subsection{General properties of \boldmath$m_{\text{eff}}$}

\label{secGenProp}One of the properties of the effective mass
$m_{\text{eff}}$ that follows immediately from its definition,
namely, its uncertainty owing to the gage invariance of the vector
potential $\mathbf{A}$, which participates in the $m_{\text{eff}}$
definition, was already noted in work \cite{Walser}.\,\,Let us
consider other properties of $m_{\text{eff}}$ that follow from the
definition\vspace*{-2mm}
\begin{equation}
m_{\text{eff}}\mathbf{v}=m\mathbf{v}+q\mathbf{A} \label{eq:9}%
\end{equation}\vspace*{-5.5mm}

\noindent but have not been noticed by anybody till now.

1. Generally speaking, the effective mass $m_{\text{eff}}$ is a
\textit{tensor} quantity.\,\,Indeed, for non-parallel $\mathbf{v}$
and $\mathbf{A}$, we also obtain that
$\mathbf{v}\nparallel\mathbf{P}$.\,\,The latter
fact means that the quantity $m_{\text{eff}}$ in the expression $m_{\text{eff}%
}\mathbf{v}=\mathbf{P}$ is a tensor rather than a scalar.\,\,The
quantity $m_{\text{eff}}$ is a scalar only in the case
$\mathbf{v}\parallel\mathbf{A}$.

An analytical expression for the tensor $m_{\text{eff}}$ can be
obtained from Eq.~(\ref{eq:9}) by regarding the vector $\mathbf{A}$
as a result of the action of some operator $\mathsf{T}$ on
$\mathbf{v}$:\vspace*{-1mm}
\[
\mathbf{A}=\mathsf{T}\mathbf{v}.
\]\vspace*{-5.5mm}

\noindent The action of this operator on the vector $\mathbf{v}$
consists in the elongation of the latter by the factor $A/v$ and the
rotation of the result until the new vector becomes superposed on
$\mathbf{A}$.\,\,Proceeding from that, let us express $\mathsf{T}$
in the form\vspace*{-1mm}
\[
\mathsf{T}=\frac{A}{v}\mathsf{R}(\mathbf{n},\varphi),
\]\vspace*{-5mm}

\noindent where $\mathsf{R}(\mathbf{n},\varphi)$ is the operator of
rotation by the angle $\varphi$ around the axis defined by the unit
vector $\mathbf{n}$, with\vspace*{-1mm}
\[
\varphi=\angle(\boldsymbol{v},\boldsymbol{A}),\quad n=\frac{\boldsymbol{v}%
\times\mathbf{A}}{\left\vert \boldsymbol{v}\times\mathbf{A}\right\vert }.
\]\vspace*{-5mm}

\noindent In such a manner, we obtain
\vspace*{-1mm}
\[
m_{\text{eff}}=m\cdot\mathbf{1}+\frac{qA}{v}\mathsf{R}(\mathbf{n},\varphi)
\]\vspace*{-5mm}

\noindent from Eq.~(\ref{eq:9}) or, in the coordinate representation,\vspace*{-1mm}
\[
\left(  m_{\text{eff}}\right)  _{ij}=m\delta_{ij}+\frac{qA}{v}R_{ij}%
(\mathbf{n},\varphi),
\]\vspace*{-5mm}

\noindent where $\mathbf{1}\equiv\{\delta_{ij}\}$ is the unit
operator ($\delta_{ij}$ is
the Kronecker symbol), the matrix elements $R_{ij}$ of the operator $\mathsf{R}%
(\mathbf{n},\varphi)$ are known to have the form (see, e.g.,
work~\cite{Biedenharn})\vspace*{-1mm}
\[
R_{ij}(\mathbf{n},\varphi)=\delta_{ij}\cos\varphi-\epsilon_{ijk}n_{k}%
\sin\varphi+n_{i}n_{j}(1-\cos\varphi),
\]\vspace*{-6mm}

\noindent and $\epsilon_{ijk}$ is the Levi-Civita symbol.

2.\,\,In the case where the effective mass $m_{\text{eff}}$ is a
scalar, it can be
both larger and smaller than par\-tic\-le's mass $m$.\,\,Moreover, $m_{\text{eff}%
}$ can accept a zero value or even a negative one.

3.\,\,In any case, the magnitude and the sign of $m_{\text{eff}}$
are determined by the relative orientation of the vectors
$\mathbf{v}$ and $\mathbf{A}$, and by the
relation between the absolute values of terms $\mathbf{p}%
=m\mathbf{v}$ and $q\mathbf{A}$, which define
$m_{\text{eff}}$.\,\,For instance, for a positively charged particle
($q>0$), we have $m_{\text{eff}}\leq0$ if the particle moves in the
direction opposite to the field $\mathbf{A}$
($\mathbf{v}\uparrow\downarrow\mathbf{A}$) and $v\leq qA/m$.

4.\,\,Generally speaking, the effective mass depends on the time: $m_{\text{eff}%
}=m_{\text{eff}}(t)$.\,\,This phenomenon is associated with the fact
that a moving charged particle passes through various spatial
points, where both the field $\mathbf{A}$ (owing to its spatial
nonuniformity or time dependence) and the particle velocity can be
\mbox{different}.

\subsection{Analogy between \boldmath$m_{\text{eff}}$ and \boldmath$m^{\ast}$}

Some of the indicated properties of Pendry's effective mass $m_{\text{eff}}$
are similar to the properties of the effective mass $m^{\ast}$ of conduction
electrons in solids.\,\,For example, in the general case, both $m_{\text{eff}%
}$ and $m^{\ast}$ can be anisotropic and should be described by the
corresponding tensors\,\footnote{Strictly speaking, as was mentioned
in Introduction, it is the inverse mass $1/m^{\ast}$ rather than the
mass $m^{\ast}$ itself that has tensor properties.}.\,\,Those
properties, in particular, distinguish both effective masses
$m_{\text{eff}}$ amd $m^{\ast}$ from the ordinary rest mass of
particles $m$, which is always a scalar larger than or equal to
zero.

\subsection{Interpretation of \boldmath$m_{\text{eff}}$}

Which is the physical meaning of the mass $m_{\text{eff}}$?\,\,Is it
\textquotedblleft real\textquotedblright?\,\,Do the giant values of
$m_{\text{eff}}$ for electrons in metal wire mesh structures (and
not only in them, see below) mean that the electron mass $m$
actually grows gigantically, so that the electron becomes extremely
heavy?

The assumption that the electron mass increases in the magnetic
field looks unphysical, because we cannot specify any real mechanism
that could be responsible for the enormous growth of the electron
mass under the physical conditions that exist in the
meshes.\,\,However, this assumption allows quantitatively proper
results to be obtained for the plasma frequency in metal mesh
structures.\,\,A collision between the nonphysical character and the
practical value of this assumption can be eliminated by giving a
proper interpretation of $m_{\text{eff}}$.

As was mentioned in Introduction, the issue concerning the physical
content of $m_{\text{eff}}$ is similar to those arisen in due time
with respect to $m^{\ast}$.\,\,Taking this fact into account, as
well as a certain similarity (not only a terminological one, see the
previous item) between $m_{\text{eff}}$ and $m^{\ast},$ we can make
attempt to interpret $m_{\text{eff}}$ in the spirit of
$m^{\ast}$.\,\,Namely, we may consider that {\it a real charged
particle with mass $m$ moves under the action of electromagnetic
forces in a magnetic field with a potential $\mathbf{A},$ as a
certain particle characterized by the same charge and the mass $m_{\text{eff}%
}$ moves under the action of the same forces but without magnetic
field.}\,\,Here, as well as in the interpretation of the effective
mass $m^{\ast}$ in the solid-state theory \cite{Tsidilkovsky}, a key
issue is that $m_{\text{eff}}$ makes it possible to calculate the
particle acceleration (in this case, in the magnetic field) under
the action of only electromagnetic forces.\,\,The acceleration of
the same particle owing to the action of forces of any other origin,
e.g., gravitational forces, depends on its rest mass $m$, which
remains the same in the magnetic field, as it was in the absence
\mbox{of a field.}\looseness=1

It is easy to see that the proposed interpretation at least agrees
with the results obtained for the metal wire mesh structures
described in Section 2.\,\,Does it contradict other known results?
Let us consider some examples.

\section{Effective Mass \boldmath$m_{\text{eff}}$ from Different Sides}

In this section, we analyze Pendry's effective mass $m_{\text{eff}}$
in various situations where charges move in a magnetic field.\,\,In
so doing, we ignore the problem associated with the ambiguity in the
choice of a field
potential $\mathbf{A}$ and, accordingly, the ambiguity of a mass $m_{\text{eff}%
}$.\,\,Similarly to works \cite{Pendry96,Ramakrishna,Cai,Pendry98},
it is supposed that the results presented below are valid for a
certain calibration of the potential $\mathbf{A}$\textbf{,} the
choice of which is not specified and remains offscreen.

\subsection{Electron in the magnetic\\ field of a conductor with current}

\label{secCurr}In Section 2, it was demonstrated that the effective
electron mass $m_{\text{eff}}$ can easily reach large (of an order
of 10$^{4}m$) values in metal mesh structures at a proper choice of
their parameters.\,\,Now, let us estimate the effective mass of an
electron moving in the magnetic field of an infinitely long
conductor with current.\,\,For a single straight infinite conductor
with radius $r_{0}$ and with current $I$, the vector potential
$\mathbf{A}$ of the magnetic field created by the current is
directed in parallel to the conductor and looks like \cite{Batygin}
\begin{equation}
A(r)=-\frac{\mu_{0}I}{4\pi}\left(\!  1+2\ln\frac{r}{r_{0}}+C
\!\right)\!\!,\label{eq:10}%
\end{equation}
where $r$ is the distance from the conductor axis, and $C$ is an
arbitrary constant.\,\,For the choice \mbox{$C=$}\linebreak
$=\mu_{0}I/(4\pi)$, potential (\ref{eq:10}) has the
\textquotedblleft minimal form\textquotedblright%
\begin{equation}
A(r)=\frac{\mu_{0}I}{2\pi}\ln\frac{r}{r_{0}}.\label{eq:11}%
\end{equation}
This formula almost coincides with expression (\ref{eq:6}) for the potential
of a magnetic field induced by currents in wire mesh structures: the
difference consists in the substitution $a\rightarrow r_{0}$.

Let us consider an electron that moves with the velocity $v=1$~mm/s
(this is a typical order of magnitude for the drift velocity of
electrons in metals at a current density of 1~A/mm$^{2}$) at the
distance $r=1$~m from the axis of a conductor with the radius
$r_{0}=1~$mm and the current $I=1$~A.\,\,According to
Eq.~(\ref{eq:11}), the magnetic field potential of the current at
the electron position equals $A\approx10^{-6}~\mathrm{T\cdot m}$.
The corresponding contribution $eA$ to the canonical momentum
(\ref{eq:2}) of the electron has an order of magnitude of
$10^{-25}~\mathrm{kg\cdot m/s}$.\,\,At the same time, the kinematic
momentum of the electron $p=mv$ under the same conditions has an
order of magnitude equal to $10^{-33}~\mathrm{kg\cdot
m/s}$.\,\,Since $eA/p\approx10^{8}$, we obtain in this case that
$m_{\text{eff}}\approx10^{8}m$.

Let us pay attention to the following issues.

1.\,\,In order to achieve the enormous values of
$m_{\text{eff}}\approx10^{8}m$, it is not necessary to use
superstrong magnetic fields or create some exotic physical
conditions.\,\,Such values can be easily realized under usual
laboratory conditions.

2.\,\,Do the large values of $m_{\text{eff}}$ mean that electrons
become extremely heavy indeed? If it have been so, then, as an
example, the mass of the rotor windings of electric motors (and the
mass of the electric motors in whole) would have increased by
several orders of magnitude at their start.\,\,Surely, it was never
observed.\,\,The interpretation proposed in item~3 (see
Subsection~3.3) allows the misleading interpretation of great
$m_{\text{eff}}$-values as the effect of an increase in the electron
mass $m$ to be avoided.

\subsection{Electron in Earth's magnetic field}

In the majority of electric engineering problems, the magnetic field
of the Earth is neglected, because the corresponding effects are
small.\,\,However, the effective electron mass $m_{\text{eff}}$ in
this field turns out even larger than in the example with the
magnetic field of the current considered in Subsection~4.1.

Really, the induction of the magnetic field created by a rectilinear
current at a distance of 1~m equals $B=\mu_{0}I/(2\pi
r)=0.2~\mu\mathrm{T}$.\,\,The magnetic field of the Earth near its
surface is known to be equal to approximately 50$~\mu\mathrm{T}$
\cite{PhysDictionary}, i.e., it is by two orders of magnitude
stronger.\,\,From the linear relation between $\mathbf{B}$
and $\mathbf{A}$ (this is a consequence of the relation $\mathbf{B}%
=\mathrm{rot}\mathbf{A}$), it follows that the potentials of those
two fields are also different by two orders of magnitude.\,\,In view
of the results of the previous example, we arrive at the conclusion
that the effective mass of an electron moving near the Earth suface
at the same velocity $v=1$~mm/s should differ from the rest mass of
electron by 10~(!) rather than 8 orders of magnitude.\vspace*{-2mm}

\subsection{Free charges in electric\\ and magnetic fields}

The analysis of free charge motions in electric and magnetic fields
of various configurations~-- uniform or non-uniform, parallel or
crossed, stationary or alternating~-- is of interest for a number of
physics domains, e.g., the physics of accelerators and the plasma
physics. Results of this analysis are well-known (see, e.g., works
\cite{Sturrock,Jackson}) and predict that, for a description of the
charge motion to be correct, it is necessary that the both
components of the Lorentz force acting on the charge~-- the
electric, $q\mathbf{E}$, and magnetic,
$q\mathbf{v}\times\mathbf{B}$, ones~-- be taken into account in the
explicit form.\,\,In contrast to that, in the framework of the
effective mass concept, the magnetic field can be excluded from
consideration, so that the motion of a particle with the mass
$m_{\text{eff}}$ only in the electric field can be analyzed (see
item~3 in Section~3.3.1).\,\,It is clear that the well-known
classical results are not reproducible in the framework of
this consideration.\,\,For instance, instead of the known $\mathbf{E}%
\times\mathbf{B}$-drift of particles in crossed $\mathbf{E}$ and
$\mathbf{B}$ fields, which is a complicated motion in the plane
perpendicular to $\mathbf{E}$ and $\mathbf{B}$ \cite{Jackson} we
obtain a simple uniformly accelerated motion of a particle with the
mass $m_{\text{eff}}$ along the \mbox{vector~$\mathbf{E}$} in the
framework of the effective mass concept.

\subsection{Influence of \boldmath$m_{\text{eff}}$ on electric conductivity}

The authors of work \cite{Solymar} paid attention to that the giant
values of effective electron mass in metal mesh structures should
induce not only a reduction of the mesh plasma frequency in
comparison with that for the mesh conductor material, but also to a
corresponding giant reduction of the mesh conductivity and a growth
of Joule losses in the mesh.\,\,It can be explained by the relation
between the conductivity $\sigma$ of the medium and the mass of
charge carriers in it.\,\,In the case of electron conductivity,
\cite{Ashcroft}\vspace*{-1mm}
\begin{equation}
\sigma=\frac{ne^{2}\tau}{m},\label{eq:12}%
\end{equation}
where $\tau$ is the electron relaxation time.\,\,For crystalline
solids, the effective mass $m^{\ast}$ characterizing the dynamic
properties of electrons in a crystal is used instead of $m$.\,\,In
the case of mesh structures, the substitution $m\rightarrow
m_{\text{eff}}$ should be made in Eq.~(\ref{eq:12}).\,\,Then,
bearing in mind the numerical estimations for $m_{\text{eff}}$, the
conductivity $\sigma$ should decrease, and, accordingly, the Joule
losses should increase by several orders of magnitude.\,\,As is
well-known, neither of those effects is observed in mesh structures.

With regard for the results described in items~1 and 2 in Section~4,
we may assert that the influence of $m_{\text{eff}}$ on the electric
conductivity should be substantial not only in mesh structures, but
also in bulk metal specimens.\,\,For example, item~2 implies that,
under the conditions of terrestrial laboratories, the conductivity
and the resistivity of an ordinary metal conductor can differ by
10~orders of magnitude from the values that would be obtained in the
absence of Earth's magnetic field.\,\,It is clear that this
prediction contradicts the whole body of accumulated experimental
data.

\subsection{Electrons in a solid}

Let us demonstrate now that the concept of effective mass
$m_{\text{eff}}$ is not pertinent to the quantum-mechanical
solid-state theory as well.\,\,First, it will be recalled that the
effective mass $m^{\ast}$ of conduction electrons used in the
solid-state theory implicitly makes allowance for the effect of
electron interaction with a periodic field $U(\mathbf{r})=U(\mathbf{r}%
+\mathbf{a})$ in the crystal lattice, so that this field can be
excluded from the Schr\"{o}dinger equation.\,\,In this case, the
Hamiltonian of the initial Schr\"{o}dinger equation,\vspace*{-1mm}
\[
\hat{H}=\frac{1}{2m}(\hat{\mathbf{p}}-e\mathbf{A})^{2}+U+V,
\]
which describes a real electron with mass $m$ that moves in the
periodic crystal field $U$, an external magnetic field with the
vector potential $\mathbf{A}$, and a field $V$ created by external
forces of all other types, e.g., electric and gravitational ones, is
reduced to a simpler form\,\footnote{Here, for simplification, the
simplest case of isotropic quadratic dispersion law is considered,
for which the effective mass is a scalar.}\vspace*{-2mm}
\begin{equation}
\hat{H}=\frac{1}{2m^{\ast}}(\hat{\mathbf{p}}-e\mathbf{A})^{2}+V,\label{eq:13}%
\end{equation}
corresponding to a free particle with mass $m^{\ast}$ moving in the fields
$\mathbf{A}$ and $V$ \cite{Davydov}.

Can the Hamiltonian of an electron in the solid be simplified
further with the use of the concept of effective electron mass
$m_{\text{eff}}$ dependent on the magnetic field? According to the
interpretation of $m_{\text{eff}}$, Hamiltonian (\ref{eq:13}) could
be simplified by excluding $\mathbf{A}$ from it and simultaneously
by making the substitution $m^{\ast}\rightarrow m_{\text{eff}}$ (in
this case, $m_{\text{eff}}$ should be calculated from
Eqs.~(\ref{eq:1}) and (\ref{eq:2}), in which the substitution
$m\rightarrow m^{\ast}$ was preliminarily made).\,\,As a result, we
obtain the following Hamiltonian of a free particle with the mass
$m_{\text{eff}}$ that moves in the field $V$:\vspace*{-1mm}
\[
\hat{H}=\frac{1}{2m_{\text{eff}}}\hat{\mathbf{p}}{}^{2}+V.
\]

Do the solutions of the Schr\"{o}dinger equation with this
Hamiltonian coincide with the solutions corresponding to Hamiltonian
(\ref{eq:13})? It is clear that this is not the case.\,\,This
becomes the most evidently if $V=0$.\,\,All the practice of
considering various phenomena in solid-state physics, in which the
magnetic field plays an essential role, e.g., the cyclotron
resonance and the Hall effect, testifies that, for those phenomena
to be described correctly, the dependence of $\hat{H}$ on
$\mathbf{A}$ must be taken into consideration explicitly.\,\,Hence,
the well-known results of the solid-state theory cannot be
reproduced in the framework of the concept of effective mass
$m_{\text{eff}}$.\vspace*{-1mm}

\section{Conclusions}

The concept of effective mass of an electron in the magnetic field
was introduced in work \cite{Pendry96}, while solving the problem of
the plasma frequency of metal mesh structures.\,\,The way of
introducing the mass $m_{\text{eff}}$ substantially uses the concept
of canonical momentum, $\mathbf{P}\equiv\partial
L/\partial\mathbf{v}$, of charged particle in an external
electromagnetic field and the relation between the kinematic,
$\mathbf{p}\equiv m\mathbf{v}$, and canonical particle's momenta.
The concept of canonical momentum is an important element of the
Lagrange and Hamilton formalisms of classical mechanics; it is also
used in the quantum mechanics.\,\,The universal character of this
notion and the relationship between $\mathbf{P}$ and $\mathbf{p}$
give a hope for that the concept of effective mass $m_{\text{eff}}$
can be used not only in the specific problem, where it was
introduced, but also in other problems (both classical and
quantum-mechanical ones) dealing with the motion of charges in the
magnetic~\mbox{field.}

The analysis of the effective mass $m_{\text{eff}}$, which was
carried out in this work, testifies to its rather unusual
properties.\,\,For instance, $m_{\text{eff}}$ turns out, generally
speaking, to be a tensor quantity.\,\,In some cases, it can acquire
zero or even negative values.\,\,A certain analogy between
$m_{\text{eff}}$ and the effective mass $m^{\ast}$ known from the
solid-state theory allows the former to be interpreted in the spirit
of the latter, by removing, in such a way to a certain extent, the
problem of a non-physical character of $m_{\text{eff}}$ marked
earlier by various authors.\,\,In addition, this interpretation of
$m_{\text{eff}}$ does not contradict the experimentally confirmed
results of work \cite{Pendry96}.

Although the relationships, on which the definition of
$m_{\text{eff}}$ is based, have universal character, the application
of the $m_{\text{eff}}$ concept beyond the problematics of the
dielectric response of metal wire mesh structures does not allow one
to obtain plausible results or to reproduce the already known
ones.\,\,For example, the dependence of the effective electron mass
on the magnetic field gives rise to a conclusion about an enormous
increase of the metal resistance even in weak magnetic fields.
Taking this fact into account, this concept has to be recognized as
an \textit{ad hoc} hypothesis, which allows the proper results to be
obtained only for a specific problem dealing with the frequency of
plasma oscillations in thin metal wire mesh
\mbox{structures.}

\vspace*{-5mm}
\rezume{%
В.В.\,Гоженко}{ПРО ЕФЕКТИВНУ МАСУ ЕЛЕКТРОНА ЗА ПЕНДРІ} {У 1996 р.
англійський фізик-теоретик Дж. Пендрі, пояснюючи діелектричний
відгук металевих сіток, висунув ідею про залежність ефективної маси
електрона від магнітного поля. Ця ідея ґрунтується на відомому
співвідношенні між кінематичним і канонічним імпульсами заряду, що
рухається в магнітному полі. В даній статті, виходячи з
універсальності зазначеного співвідношення, досліджується можливість
застосування поняття ефективної маси електрона за Пендрі $m_{\rm
eff}$ не лише до електронів у металевих сітках, а й до більш
широкого кола задач про рух зарядів у магнітному полі. Встановлено
загальні властивості ефективної маси $m_{\rm eff}$, які випливають
безпосередньо з її означення. Виявлено аналогію між $m_{\rm eff}$ та
ефективною масою електронів $m^{*}$, що розглядається у теорії
твердого тіла. Запропоновано фізичну інтерпретацію $m_{\rm eff}$. На
кількох прикладах показано, що, незважаючи на універсальність
використовуваного в означенні $m_{\rm eff}$ співвідношення між
кінематичним і канонічним імпульсами, застосування поняття $m_{\rm
eff}$ поза межами проблематики діелектричного відгуку металевих
сіток не дозволяє одержати правильні результати.}

\end{document}